\documentclass[11pt,twoside]{article}
\usepackage{asp2010}

\resetcounters

\begin{document}

\title{Dust at sub-solar metallicity: the case of post-AGB stars in the Large Magellanic Cloud}
\author{M.~Matsuura$^{1,2}$, G.~C.~Sloan$^3$, J. Bernard-Salas$^{3,4}$,  K.~Volk$^5$, and B.~J.~Hrivnak$^6$
\affil{$^1$ UCL-Institute of Origins, Department of Physics and Astronomy, University College London, 
	Gower Street, London WC1E 6BT, UK}
\affil{$^2$UCL-Institute of Origins, Mullard Space Science Laboratory,
University College London, Holmbury St. Mary, Dorking, Surrey RH5 6NT, UK}
\affil{$^3$Astronomy Department, Cornell University, 610 Space Sciences Building, 
        Ithaca, NY 14853-6801, USA }
\affil{$^4$Institut dÕ Astrophysique Spatiale, CNRS/Universite Paris-Sud 11, 91405 Orsay, France}
\affil{$^5$Space Telescope Science Institute, 3700 San Martin Drive, Baltimore, MD 21218, USA}
\affil{$^6$Department of Physics and Astronomy, Valparaiso University, Valparaiso, IN 46383, USA}}

\begin{abstract}
 Low- and intermediate-mass stars are one of the important dust sources in the interstellar medium (ISM) of galaxies.
The compositions of dust ejected from these stars are likely to affect those in the ISM.
We investigate dust in post-Asymptotic Giant Branch (AGB) stars, which are in a late evolutionary phase
for low- and intermediate-mass stars, and which produce a wide variety of dust grains.
We are particularly targeting post-AGB stars in the Large Magellanic Cloud (LMC), which has about half of the solar
metallicity, to investigate the effects of sub-solar metallicity on dust compositions.

 Using the Spitzer Space Telescope, we obtained 5--30\,$\mu$m spectra of 24 post-AGB candidates in the LMC.
Five are C-rich post-AGB stars, and this presentation focuses on spectra of these stars.
 
 We found that rare dust features in the Milky Way, such as a 21\,$\mu$m unidentified feature are commonly found in LMC  post-AGB stars.
The 6--8\,$\mu$m spectra are compared with those of Galactic objects. 
Four spectra match the Galactic templates of polycyclic aromatic hydrocarbon (PAH) features. 
However, we found the three objects show 7.85\,$\mu$m feature which have not found in Galactic post-AGB stars.
Low metallicity conditions definitely affect the dust formation process and compositions.

\end{abstract}

\section{Introduction}

   There are four key aspects regarding dust grains to be investigated
in AGB stars, post-AGB stars and planetary nebulae (PNe).
First, the dust grains maintain the thermal balance within the nebulae, 
absorbing UV and optical photons, and sometimes producing large infrared excesses \citep{Kwok00}.
Second, dust grains are an important catalyst for molecular formation.
In particular, chemical reactions on grains are required to produce molecular hydrogen \citep{Hollenbach71},
including in PNe \citep{Aleman04, vanHoof10}.
Third, dusty disks and tori might trigger asymmetry in post-AGB stars and PNe.
Finally, these evolved stars are one of the important contributors of dust found in the interstellar medium (ISM)
of galaxies \citep{Gehrz89, Matsuura09}.  
The importance of dust is reported in the White Paper developed during this conference.

Our work is specifically directed to comparing dust found in circumstellar envelopes (CSEs) of AGB/post-AGB/PNe to the dust in the ISM.
As such, we will be able to reveal
if dust has been subsequently processed in the ISM. 
Our main focus is the dust in post-AGB stars,
 where wide varieties of dust species have been detected.

  Our targets are post-AGB stars in  the Large Magellanic Cloud (LMC), our neighbouring galaxy. 
One of the advantages is that the distance of post-AGB stars in this galaxy
are independently measured, whereas the distance is an unknown parameter for
the Galactic post-AGB stars. 
This is ideal since it allows one to trace the stellar evolution on an HR diagram.
Secondly, the LMC provides a laboratory
at sub-solar metallicity. Dust grains contain metals and so
the mass and composition of dust might change according to the metallicity.
Finally, the close distance is favourable for resolving the constituent stars in the galaxy,
and recent advances in observational techniques in the infrared enable us to 
carry out detailed studies of objects belonging to this galaxy
for the first time.

 In this paper, we report our observational studies of dust in the LMC,
 indicating the effects of metallicity on the dust grains in the post-AGB stars.

\section{Observations}

   We have obtained spectra of  24 post-AGB candidates in the LMC, using the Infrared Spectrometer \citep[IRS; ][]{Houck04}
on board the Spitzer Space Telescope.
We used the low-spectral resolution mode for the majority of targets. 
The spectral resolution was $\lambda/\Delta\lambda$=60--130.
We have chosen the targets largely based on infrared brightness, using the Spitzer imaging survey data
of the LMC  \citep[SAGE; ][]{Meixner06}. 
 \citet{Matsuura11} describe the details of the sample selection and provide an observing log.

\begin{figure} 
    \resizebox{\hsize}{!}{\includegraphics[angle=90]{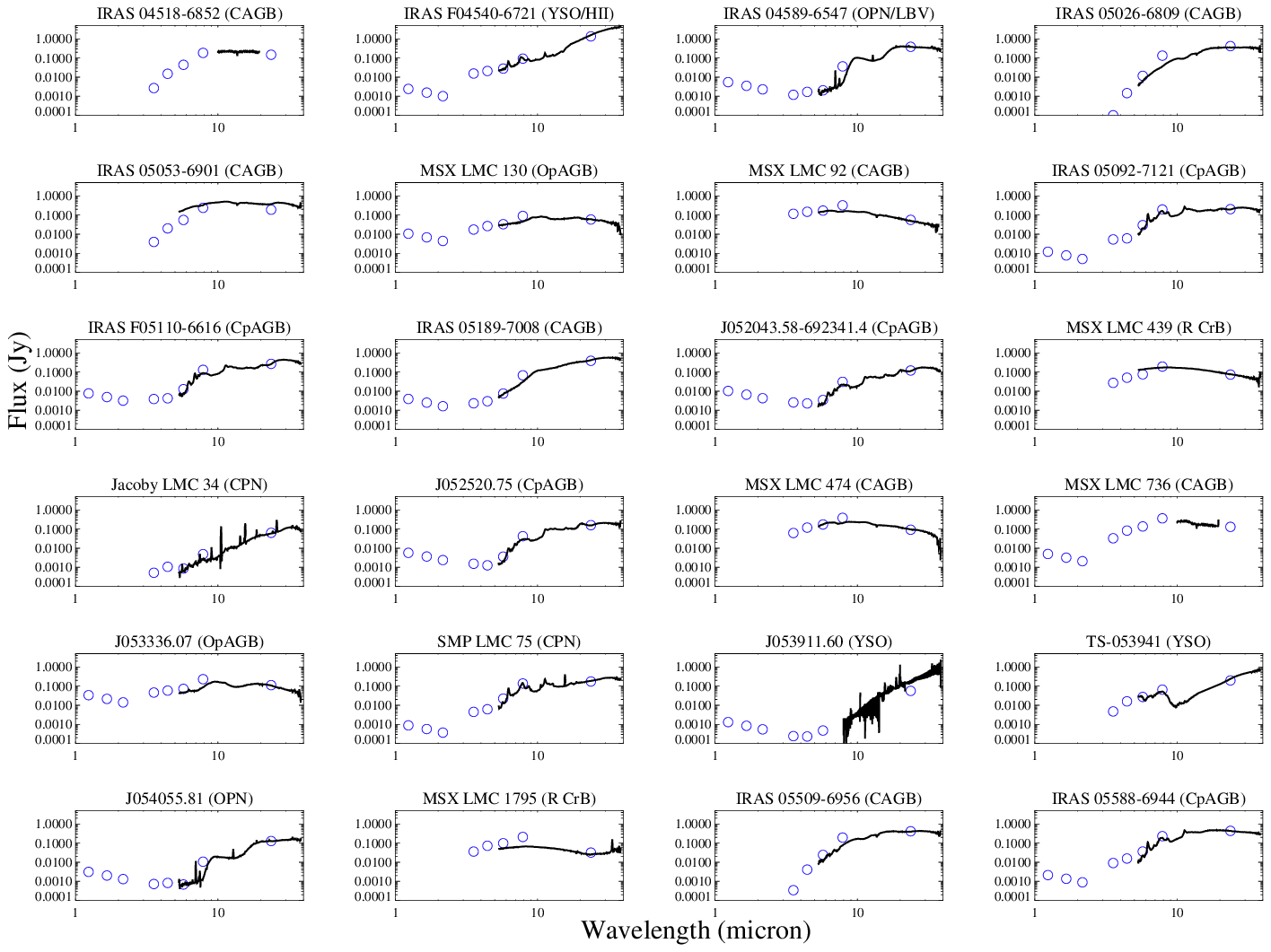}} 
   \caption{Spitzer IRS spectra of LMC IR bright objects, overlaid on photometric data
   at 1--24\,$\mu$m. Object classifications show 
   CAGB (C-rich AGB), 
   CPN (C-rich PN),
   CpAGB (C-rich post-AGB),
   LBV (luminous blue variable),
   OPN (O-rich PN), 
   OpAGB (O-rich post-AGB),
   R CrB,
   YSO (young stellar object).
   \label{sed}}
\end{figure}

   Mid-infrared spectroscopy is a powerful tool to distinguish object types 
(AGB, post-AGB, PNe, and R CrB)
of infrared-bright evolved stars, as well as chemical types
(oxygen-rich or carbon-rich; no obvious S-type stars found).
Figure\,\ref{sed} shows the spectra of the 24 targets we observed.
We describe the characteristics and categorise post-AGB objects 
in our paper; we found 5 carbon-rich (hereafter, C-rich) post-AGB stars and 2 oxygen-rich (O-rich) post-AGB stars.
Other types we found among our sample were 8 C-rich AGB stars, 2 R CrB candidates,
3 C-rich PNe, 3 young stellar objects (YSOs)
and 1 luminous blue variable (LBV).
Among our sample, there are no O-rich AGB stars.

  The largest category in our sample is C-rich AGB stars.
The stars we detected are special amongst C-rich AGB stars,
the so-called `extremely red objects' (EROs)  \citep{Gruendl08}.
The spectral energy distributions peak in the 10--20\,$\mu$m range,
often with a SiC dust feature at 11.3\,$\mu$m in absorption. 
C-rich stars commonly show C$_2$H$_2$ in absorption \citep{Matsuura06}. 
EROs still show C$_2$H$_2$  in absorption, but very weakly. 
Some of them (e.g. IRAS\,05189$-$7008 in Figure\,\ref{sed}) have a secondary peak in the optical,
showing that the central star has partially cleared out its optically-thick circumstellar envelope.
It is likely that these stars are transiting from the AGB to the post-AGB phase.

\citet{Gielen09} and \citet{vanAarle09} have studied post-AGB stars in the LMC,
and their targets are mostly binary objects. They found more to be O-rich 
than C-rich. Our targets are not selected based on suspected binarity.
Although the samples are small, it is possible that binarity and chemical types may be related.

\section{Dust features in C-rich post-AGB stars}

\begin{figure} 
\centering
  \resizebox{0.45\hsize}{!}{\includegraphics*{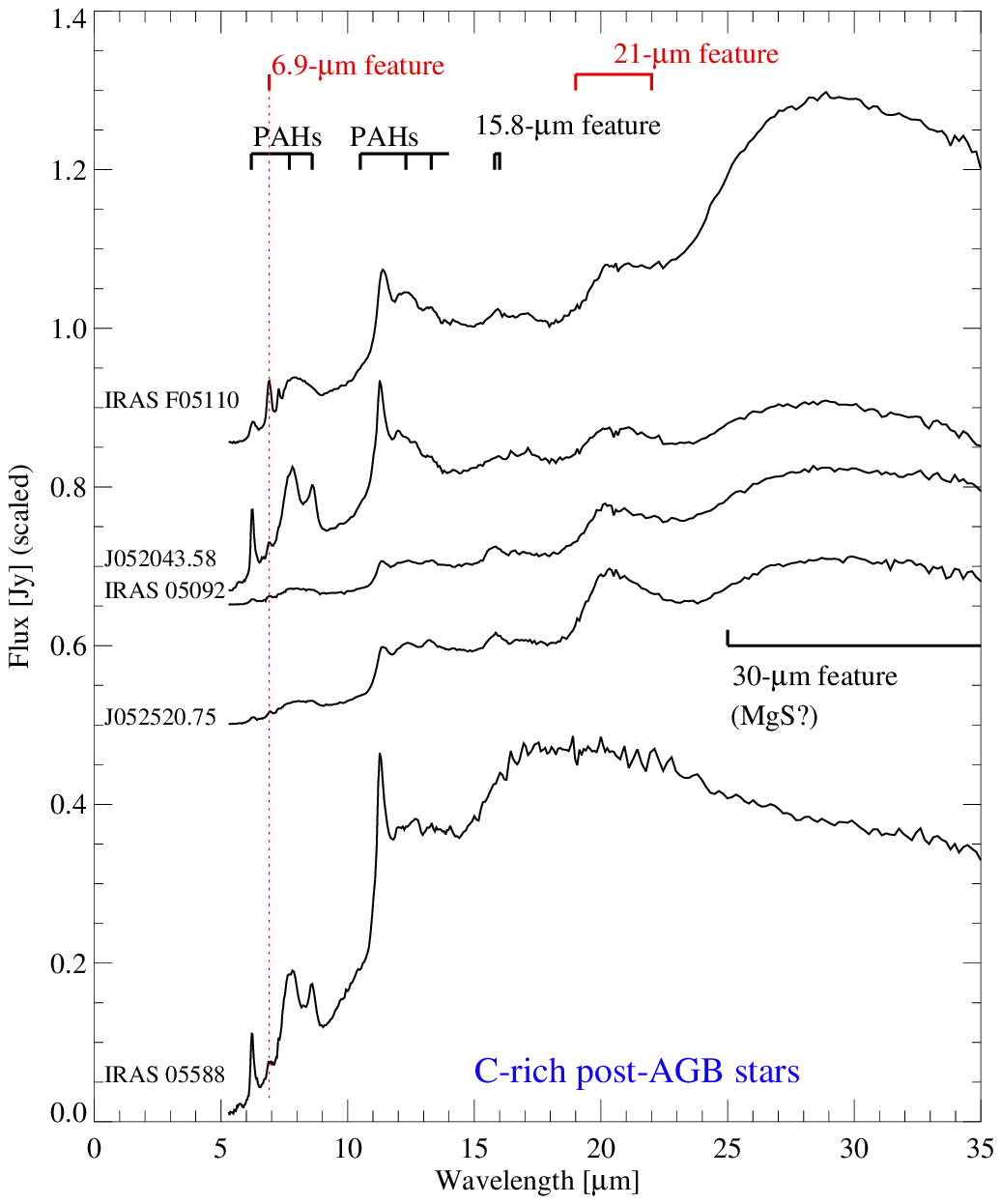}} 
   \caption{Spectra of C-rich post-AGB stars in the LMC.
   \label{spectra}}
\end{figure}

Figure\,\ref{spectra} present the Spitzer {\it IRS} spectra of the five C-rich post-AGB
stars, showing a variety of emission features.
The next section discusses the features related to PAHs, which dominate. 
The 30\,$\mu$m feature probably arises from MgS \citep{Goebel85}.
The features at 6.9\,$\mu$m and 7.3\,$\mu$m are likely to be from
aliphatic carbon \citep{Duley81, Kwok01}, although there might
be some contributions of PAHs to the 6.9\,$\mu$m feature \citep{Tielens08}.

The spectra in Figure\,\ref{spectra} also show a feature centred at 
15.8\,$\mu$m (e.g. IRAS 05092 and J052520.70) and the unidentified `21 $\mu$m' feature,
which actually peaks at 20\,$\mu$m in these spectra.
The carrier of  21-$\mu$m feature remains a point of controversy \citep[e.g.][]{Volk99, Hrivnak09}.
The details of this feature in our sample will be presented by \citet{Volk11}.
The carrier of the 15.8\,$\mu$m and 21 $\mu$m features
may be related \citep{Hrivnak09, Zhang10}.

\section{Polycyclic Aromatic Hydrocarbons (PAHs) and other features at 6--9\,$\mu$m}

\begin{figure} 
\centering
  \includegraphics[scale=0.5]{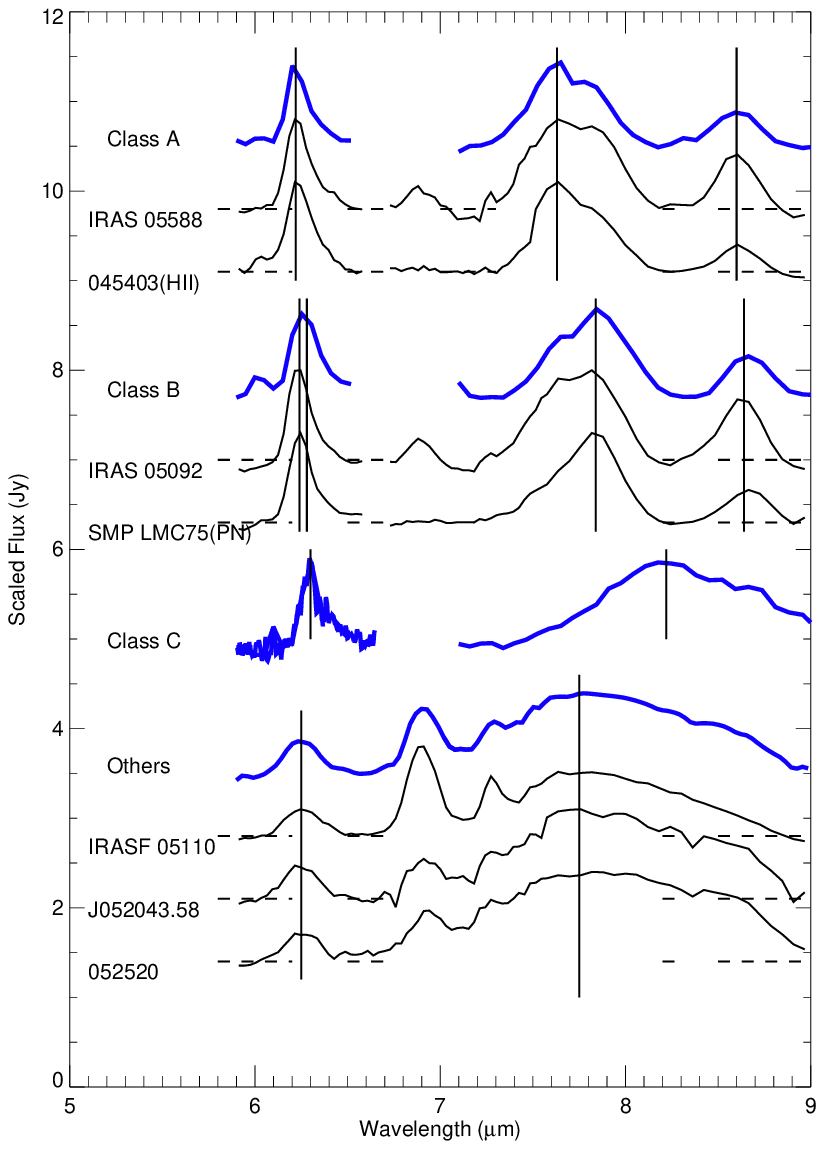}
   \caption{The 6--9\,$\mu$m spectra in LMC post-AGB stars. As references,
   PAH spectra of an HII region and planetary nebula in the LMC are plotted.
   \citet{Peeters02} and \citet{Hony02} have analysed PAHs in Galactic objects
   and classified them into three classes, A, B and C, and we plot these templates.
   The spectra of the three LMC objects match any of the Peeters classes.
   These three are called 'others'.
   \label{PAHs}}
\end{figure}

 To examine the details of PAH features and other small features, we subtract  `continua' 
from the observed spectra. 
\citet{Matsuura11} provide details of this process.
Figure\,\ref{PAHs} plots the continuum-subtracted spectra.
\citet{Peeters02} and \citet{Hony02} have analysed PAHs in Galactic objects, including post-AGB stars,
PNe, and HII regions, and these authors classified them into classes, A, B and C.
Figure\,\ref{PAHs} shows their template spectra.
The classes can be distinguished by the number of peaks and the peak wavelength, as indicated in the figure.
Both classes A and B show features peaking at about 7.65 and 7.85\,$\mu$m,
with the 7.65\,$\mu$m peak dominating in class A and the 7.85\,$\mu$m peak dominating in class B.
In class C, only a single peak appears at about 8.3\,$\mu$m.

Among our sample, two spectra can be classified into A, and another two into B. 
Three further objects show similar profiles to class C, as they have a broad single feature between 7 and 9\,$\mu$m.
However, their peak wavelengths are shorter than those found in class C.
These three spectra do not fit within the current A-B-C classification system. 
We label these three as `others', having two peaks at about 6.3 and 7.85\,$\mu$m.

In the spectra of these three `others', the  6.3\,$\mu$m feature is due to PAHs.
However, the feature of these three others is broader than normally observed, and it much more symmetric.
Overall, the behaviour of this feature is analogous to the behaviour of 7-9\,$\mu$m feature,
and together these suggest a carrier that differs in some way from a normal PAH mixture.

The 7.7--7.8 $\mu$m feature could arise from PAHs \citep{Sandford98a, Sandford98b}
or the vibrational transition of -(CH$_2$)$_n$- in
aliphatic carbon \citep{Socrates01}.  
However, the width of the
feature probably requires multiple components rather than a
solitary aliphatic-carbon band.


Although Galactic class C and LMC the 7.85 $\mu$m feature objects have central stars with similar spectral types 
\citep[F and G; ][]{Peeters02, Sloan07},
the features appear at slightly different wavelengths.
It could be due to the difference of PAH compositions between LMC and Galactic objects.
Whereas in class A and B PAHs, there is no difference between LMC and Galactic objects.
The difference is found only in stars with spectral classes of F and G-type.


 Among C-rich post-AGB stars, we did not find any difference in the strength 
 of PAHs between Galactic and LMC objects.
  This is consistent with what \citet{Bernard-Salas09}
 found in PNe. 
  Among objects with 21\,$\mu$m features, which
 are a sub-class of post-AGB stars, PAHs are stronger \citep{Volk11}.
At  lower metallicity, PAHs could be formed reasonably well.
This is because C$_2$H$_2$, which is a parent molecule for the PAH formation
\citep{Allamandola89},
 is abundantly found in AGB stars in lower metallicity galaxies \citep{Matsuura05a, Matsuura07, Sloan09}
since carbon atoms are synthesised within these stars
and since the oxygen abundance is low, resulting in a high excess of available carbon.

 We acknowledge the financial support of NASA grant (JPL-RSA 1378453).
 
\bibliography{aspauthor}

\end{document}